\pdfoutput=1
\documentclass[aps,pre,twocolumn,groupedaddress,amssymb,amsmath,nofootinbib]{revtex4}

\usepackage{amssymb,amsmath}
\usepackage{graphicx}
\usepackage{subfigure}
\newcommand{\beq}{\begin{equation}}
\newcommand{\eeq}{\end{equation}}
\newcommand{\beqn}{\begin{eqnarray}}
\newcommand{\eeqn}{\end{eqnarray}}
\newcommand{\rmd}{\mathrm{d}}
\newcommand{\nn}{\nonumber}
\renewcommand{\log}{\ln}

\begin{document}

\title{Interference in disordered systems: A particle in a 
complex random landscape}

\author{Alexander Dobrinevski}
\email[]{dobrinev@lpt.ens.fr}
\author{Pierre Le Doussal}
\email[]{ledou@lpt.ens.fr}
\author{Kay J\"org Wiese}
\email[]{wiese@lpt.ens.fr}
\affiliation{CNRS-Laboratoire de Physique Th\'eorique de l'Ecole
Normale Sup\'erieure, 24 rue Lhomond, 75005 Paris
Cedex-France
\thanks{LPTENS is a Unit\'e Propre du C.N.R.S.
associ\'ee \`a l'Ecole Normale Sup\'erieure et \`a l'Universit\'e Paris Sud}
}

\date{\today\ 
}

\begin{abstract}
We consider a particle in one dimension submitted to amplitude and phase disorder.
It can be mapped onto the complex Burgers equation, and provides a toy model for 
problems with interplay of interferences and disorder, such as the NSS model of
hopping conductivity in disordered insulators and the Chalker-Coddington
model for the (spin) quantum Hall effect. The model has three distinct phases: (I) a {\em
high-temperature} or weak disorder
phase, (II) a {\em pinned} phase for strong amplitude disorder, and (III) a
{\em diffusive} phase  for strong phase disorder, but weak amplitude disorder. 
We compute analytically the renormalized disorder correlator,
equivalent to the Burgers velocity-velocity correlator at long times. In phase III, it assumes a universal form. For strong phase disorder, interference leads to a logarithmic singularity, related to zeroes of the partition sum, or  poles of the complex Burgers velocity field.
These results are valuable in the search for the adequate field theory for higher-dimensional systems.
\end{abstract}


\maketitle

\section{Introduction}
Much progress has been accomplished in the understanding of the 
thermodynamics of classical disordered systems \cite{BinderYoung1986,BookYoung}.
Typically, the disorder is modeled by a random potential. At low temperature, the low lying local minima of the resulting, rough, energy landscape become metastable states, and dominate the partition sum of the system. The correlations of the random potential 
determine the statistics of these metastable states, and hence the physics of the model. In many cases, for example
in elastic disordered systems, the scaling of observables can be described by a family of $T=0$ fixed points of the RG flow (most notably random-field and random-bond), which yield characteristic, universal roughness exponents and effective disorder correlators \cite{Fisher1986a,Balents1993,LeDoussal2004}.

\begin{figure}[b]
\centering \includegraphics[width=.8\columnwidth]{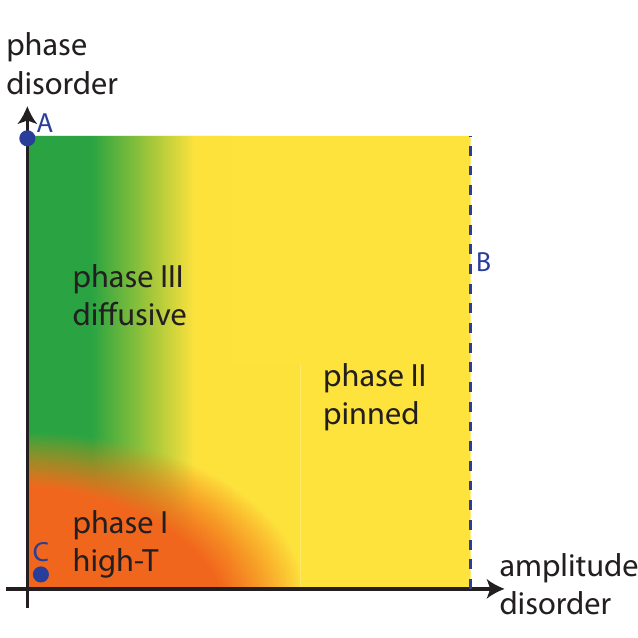}
\caption{(Color online) Phase portrait of the model. The horizontal
axis is the strength of $V$ and the vertical axis the strength of
$\theta$. The effective disorder correlators for points A (deep in the
diffusive phase), B (deep in the
pinned phase), C
(infinitesimally small disorder) are analyzed in
sections \ref{sec:Ph3}, \ref{sec:Ph2} and \ref{sec:Ph1}, respectively.}
\label{fig:CxPhases}
\end{figure}
However, the picture is much less clear when quantum interference is
important. In real time dynamics one must study 
a sum over Feynman paths, whose weights are complex random numbers. The dominant contributions may then be difficult to discern.

To give a specific example, hopping conductivity of electrons in
disordered insulators in the strongly localized regime is described by the Nguyen-Spivak-Shklovskii
(NSS) model \cite{Nguyen1985}. The probability amplitude $J(a,b)$ 
 is  the sum over interfering directed paths $\Gamma$ from  $a$ to $b$ \cite{Medina1991,Medina1992,Medina1990,Somoza2007,Prior2009}
\begin{equation}
\label{eq:IntroNSS}
J(a,b) := \sum_{\Gamma} \prod_{j\in \Gamma} \eta_j \ .
\end{equation}The conductivity between sites $a$ and $b$ (e.g.\ on a $\mathbb{Z}^d$ lattice) is then given by $g(a,b)\sim|J(a,b)|^2$. Each lattice site $j$ contributes a random sign $\eta_j=\pm 1$ (or, more generally a complex phase $\eta_j=e^{i\theta_j}$). 

Another example is the Chalker-Coddington model \cite{Chalker1988} for the quantum Hall
(and spin quantum Hall) effect, where the
transmission matrix $T$, and the conductance $g (a,b)\sim
\text{tr}\,T (a,b)^{\dagger}T (a,b)$, between
two contacts $a$ and $b$ is given by
\cite{Cardy2010,Beamond2002}: 
\begin{equation} 
\label{eq:IntroCC}
T (a,b) =\sum_{\Gamma} \prod_{(ij)\in \Gamma} S_{(ij)}\ .
\end{equation}
The  random variables $S_{(ij)}$ on every bond $(ij)$ are  $U(N)$
matrices,  with
$N=1$ for the charge quantum Hall effect and $N=2$ for the spin
quantum Hall effect. Here $\Gamma$ are paths subject to some rules imposed at the vertices. 

In both models, one would like to understand the dominating
contributions to the sum over  paths with random weights, given by
$J(a,b)$ or $T (a,b)$, respectively; we shall denote it by $Z$ in the
following. In contrast to the thermodynamics of classical models,
where all contributions are positive, contributions between paths with
different phases can now cancel. One is also interested in the
expected phase transitions, i.e.\ critical values for the amplitude
and phase disorder above which  interference effects become important.

This is a  complicated problem. In this article, we therefore
consider a toy model motivated by the models above,
for which many computations can be done explicitly. 
More precisely, we analyze a ``partition sum'' $Z$ defined in one dimension, of the form 
\begin{equation}
\label{eq:IntroZ}
Z (w) =  \sqrt{\frac{\beta m^2}{2\pi}} \int_{-\infty}^{\infty}\rmd x\, e^{-\beta \big[V(x)+\frac{m^2}{2}(x-w)^{2}\big] -i \theta(x)}
\end{equation}
Here, $V(x)$ is a random potential and $\theta(x)$ a random
phase, both with translationally invariant correlations, 
and $\beta=1/T$ the inverse temperature. This is a toy model defined in one dimension and thus a drastic
simplification compared to both the NSS model \eqref{eq:IntroNSS} and
the Chalker-Coddington model \eqref{eq:IntroCC} (which are usually
considered in two dimensions and above). However, a similar
toy model for a  random potential without random phases
 reproduces many important physical features of realistic,
higher-dimensional models. For example, the appearance of shocks and
 depinning are already present in this
framework \cite{BalentsBouchaudMezard1996,LeDoussal2006,LeDoussal2009}. For purely
imaginary disorder, a nice experimental realization of the sum in \eqref{eq:IntroZ}
in cold atom physics is discussed in Section \ref{sec:ColdAtoms}. Complex sums similar to \eqref{eq:IntroZ} are also encountered in magnetization relaxation in random magnetic fields, see e.g. \cite{Mitra1991}.
 
The interplay between random phases $\theta(x)$ and a random potential
$V(x)$  similar to \eqref{eq:IntroZ} was already studied, among
others, in \cite{Cook1990,Derrida1991,Derrida1993,Goldschmidt1992}. The basic
distinction of three phases, high-temperature phase I, frozen phase
II, and strong-interference phase III, was established by Derrida in
\cite{Derrida1991a}, 
and we follow his conventions. The aim of this paper is to pursue a
complementary approach to \cite{Derrida1991a}, based on the study of
renormalized disorder correlation functions.  The latter is
defined due to the presence of the parabolic well centered at $x=w$
in \eqref{eq:IntroZ}. The resulting spatial structure exhibits non-trivial
features such as, in some cases, discontinuous jumps (shocks) as $w$ is varied.
Furthermore the renormalized disorder
correlator is the central object of the field theoretic treatment of
disordered systems based on the functional renormalization group
\cite{LeDoussal2006,WieseLeDoussal2006}, and thus the results of the toy model will give hints for a treatment of more realistic, higher-dimensional systems.

\medskip

This article is organized as follows: In section \ref{sec:Prelim}, we
give the theoretical framework for our treatment. The model (\ref{eq:IntroZ})
is related to a complex Burgers equation, with time
$t=m^{-2}$, which has generated interest in the mathematics community recently
\cite{Kenyon2007}. Equal-time correlation functions of the Burgers velocity
field are the \textit{renormalized disorder correlation functions}
$\Delta (w-w')$ of our model, $\overline{\partial_{w} \ln (Z (w))
\partial_{w'} \ln (Z (w')) } $. The precise definition is given in
Sec.\ \ref{s:eff-dis-cor}.  They encode physical properties of the system
like the appearance of shocks.  Their $m\rightarrow 0$
(i.e.\ $t \rightarrow \infty$ in the Burgers picture) asymptotics forms the basis for the following
analysis of the various phases. 

We  first discuss the strong-interference regime, $V(x)=0$ and
sufficiently strong $\theta(x)$, in section \ref{sec:Ph3}. This is the
regime most directly related to the NSS model and the
Chalker-Coddington model described above. Naively, one may think in
analogy to the case of classical disordered systems where
$\theta(x)=0$, that points of stationary phase  take on the role of
the local minima of the energy landscape, and dominate the partition
sum. We will show that this is incorrect. Instead, fluctuations of $Z
(w)$ along the entire system are important. In our analytical
treatment using the  replica formalism, this manifests itself as a
pairing of  replicas. We will see that there is a finite density of
zeroes of $Z (w)$ (as already observed in \cite{Derrida1991a}) which
manifests itself in a logarithmic singularity of the effective
disorder correlator $\Delta (w-w')$ for $w$ close to $w'$. 

In section \ref{sec:Ph2}, we  consider the influence of random phases in the frozen regime (large $\beta V(x)$), where only a few local minima of the random potential contribute to $Z (w)$. In the $\beta\rightarrow \infty$ limit one finds sharp jumps between these minima as $w$ is varied. In the Burgers velocity profile, these manifest themselves as shocks, and their statistics are known to be encoded in a linear cusp of the effective disorder correlator \cite{BalentsBouchaudMezard1996,LeDoussal2006}.
We then discuss how the form of these shocks is modified by the
introduction of random phases. It  turns out that the linear cusp of
the effective disorder correlator again acquires a  logarithmic
singularity. This time, however, it is related to shocks between
two minima where the phase angle difference is $\pi$, i.e.\ $Z (w)$ passes through $0$. This phenomenon is dependent on the spatial structure and on the possibility to vary $w$, and hence was not observed in \cite{Derrida1991a}.

In section \ref{sec:Ph1}, for completeness we briefly discuss the
high-temperature phase.  Here, fluctuations of $Z$ are small compared
to its expectation value, and $Z$ never becomes zero. As a
consequence, the effective disorder correlators are  regular everywhere, indicating that no shocks or poles occur in the Burgers velocity field.

At any finite system size $L\sim \frac{1}{m}$, there are blurred cross-overs between these phases as shown in figure \ref{fig:CxPhases}. They become sharp transitions in the thermodynamic limit if the variance of $\theta$ and $V$ is rescaled with the system size $L$ as $\overline{V^2}\sim\overline{\theta^2}\sim \log{L}$ \cite{Derrida1991a}. Since we are interested in the behaviour of the disorder correlators deep inside each individual phase, we do not follow this path but instead choose the simpler scaling $\sim 1$ or $\sim L$. By doing this for $V$ and $\theta$ individually, we can shrink all phases but one in the phase diagram to points respectively lines, and  discuss each phase individually.
 
In conclusion, one significant physical result of our work is that the
introduction of random phases has quite different effects depending on
the real potential $V(x)$. If $V (x)$ is sufficiently strong so that
the system is in the frozen phase, even weak random-phase disorder
immediately introduces zeroes of $Z (w)$ or turns the shocks of the
real Burgers velocity profile into poles of the complex Burgers
velocity profile. In the high-temperature phase, where $V (x)$ is
weak, this does not happen for weak random phase disorder, and the effective disorder correlators remain analytic.

\section{Preliminaries\label{sec:Prelim}}
\subsection{Definition of the model}
To completely define the model \eqref{eq:IntroZ}, one needs to specify
the joint distribution of the random potential $V(x)$ and the random
phases $\theta(x)$. For the purpose of this paper, we assume
that $V$ and $\theta$ are independent, and that the distribution of
$\theta$ is symmetric around $0$. This is mostly for technical reasons
(since this choice makes many observables real) and is certainly true
e.g.\ for centered Gaussian distributions.

Typically, one chooses $V(x)$ to be Gaussian with mean zero,
$\overline{V (x)}=0$, and variance depending on the type of
correlations. 
Here $\overline{\cdots}$ denotes averages over realizations of the disorder. 
In the absence of imaginary disorder, a short-range correlated $V(x)$, 
i.e.\ $\overline{V(x)V(x')}=0$ unless $x=x'$, gives the so-called
Kida model \cite{Kida1979,Bouchaud1997,LeDoussal2009}.
If one chooses $V(x)$ to be
long-range correlated as a random walk, $\overline{[V(x)-V(x')]^{2}}\sim
|x-x'|$, one obtains the Sinai model \cite{Sinai1982}. For $\theta(x)$ we 
also assume translationally invariant correlations. 

For some computations, it is easier to regularise by a finite system size $L$,
\begin{equation}
\label{eq:CxPartSumL}
Z_{L} := \frac{1}{L}\int_0^L e^{-\beta V(x) - i \theta(x)}\, \rmd x\ .
\end{equation}
The system size $L$ can be related to the mass of the harmonic well through $L\sim \frac{1}{m}$.
In the case of pure random-phase disorder, i.e.\ $V(x)=0$,
\eqref{eq:CxPartSumL} can be seen  as a partition sum of a particle in the real random potential $\theta(x)$ at imaginary inverse temperature $i\beta$.

\subsection{Proposed measurement of $Z$ in cold atoms\label{sec:ColdAtoms}}
A direct measurement of the partition sum as given in \eqref{eq:IntroZ} for the strong-interference phase, i.e.\ with random phases $\theta(x)$ but without a random potential $V(x)$, is at least in principle possible in a cold-atom experiment: Prepare the system in the ground state of a weak 
harmonic well (with frequency $\omega$), so that at $t=0$ the wavefunction is 
\begin{equation}
\psi_0(x):=\psi(x,t=0) = \left( \frac{M\omega}{\pi\hbar}\right)^{\frac{1}{4}} e^{-\frac{M \omega x^2}{2\hbar}}\ .
\end{equation}
Then, switch off the harmonic well, and instead switch on a random potential $\theta(x)$. In situations where 
the kinetic term in the Hamiltonian is negligible, such as $\omega t \ll 1$, or a large mass $M$, the time evolution is approximately given by $e^{-\frac{i}{\hbar}\theta(x)t}$, i.e.
\begin{equation}
\psi(x,t) = \left( \frac{M \omega}{\pi\hbar}\right)^{\frac{1}{4}} e^{-\frac{M \omega x^2}{2\hbar}-\frac{i}{\hbar}\theta(x)t}\ .
\end{equation}
Switching the potential back to the harmonic well and measuring the overlap with the ground state gives
\begin{equation}
\left< \psi_0 | \psi(t) \right> = \left( \frac{M \omega}{\pi\hbar}\right)^{\frac{1}{2}} \int_{-\infty}^{\infty}\rmd x \, e^{-\frac{M\omega x^2}{\hbar}-\frac{i}{\hbar}\theta(x)t}\ .
\end{equation}
This is exactly of the form of $Z(w)$ in \eqref{eq:IntroZ}. Although the overlap $\left\langle \psi_0 | \psi(t) \right\rangle$ cannot be measured directly, the occupation probability of the ground state, given by $|\langle \psi_0 | \psi(t)\rangle|^2$,  could in principle be measured, providing a direct measurement of $|Z(0)|^2$ in the strong-interference regime.
An example of a related experiment is given in \cite{Ovchinnikov1999}.

\subsection{Complex Burgers equation\label{sec:CxBurgers}}
Another application of the toy model \eqref{eq:IntroZ} is to the complex Burgers equation.
With the mapping $t:=m^{-2}$,  one obtains from \eqref{eq:IntroZ} the
``equation of motion'' or ``renormalisation group equation'' ($m^{2}$
being interpreted as the infrared cutoff) for $Z$,
\begin{equation}\label{9}
\partial_{t} Z (w,t) = \frac{T}{2} \partial_{w}^{2} Z (w,t)\ .
\end{equation}
We have added the argument $t$ for clarity. The initial condition at
$t=0$ equivalent to $m^2=\infty$ is 
\begin{equation}\label{9b}
Z (w,0) = e^{-\beta V(w)-i \theta(w)}\ .
\end{equation}
Using the Cole-Hopf transformation $h (w,t) := -T \ln Z (w,t)$, we obtain the
KPZ-equation 
\begin{equation}\label{10}
\partial_{t} h (w,t) = \frac{T}{2} \partial_{w}^{2}h (w,t)
-\frac{1}{2} \left[\partial_{w}h(w,t) \right]^{2} \ .
\end{equation}
Taking one spatial derivative one arrives at the Burgers equation for the velocity $u(w,t):=\partial_{w} h(w,t)$:
\begin{equation}
\label{eq:CxBurgers}
\partial_t u(w,t)=\frac{T}{2}\partial_w^2 u(w,t) - u \partial_w u(w,t)\ .
\end{equation}
Without random phases ($\theta(x)=0$), $u(w,t)$ is real and
\eqref{eq:CxBurgers} is the well-studied {\em real} Burgers
equation. It has been used, among others, to describe the formation of
large-scale structures in cosmology (the so-called \textit{adhesion
model}) and in compressible fluid dynamics (for a review see \cite{Bec2007}). 
When random phases are included, $u$ becomes complex. The resulting complex Burgers equation has very surprising applications e.g. to Lozenge tilings of polygons \cite{Kenyon2007}. It has also been used to obtain further information on the real Burgers equation through analytic continuation to the complex plane and the so-called \textit{pole expansion} \cite{Senouf1997,Senouf1996,Bessis1984,Bessis1990}.
In the following, we study the small-$m$ properties of
\eqref{eq:IntroZ}, or equivalently the large-$t$ properties of the complex Burgers equation \eqref{eq:CxBurgers}. 

\subsection{Effective disorder correlators}\label{s:eff-dis-cor}
The main observables on which we base our analysis are the
so-called \textit{effective disorder correlators}, which we define now.
For each realization of the random potential $V(x)$ and the random phases $\theta(x)$, we first define the ``free energy'' or the ``effective potential''  by
\begin{equation}\label{eq:DefEffPot}
\beta \hat{V}(w) + i\hat{\theta}(w) := \beta h (w)\equiv  -\ln Z (w) \ .
\end{equation}
Note that $\hat{V}(w)$ is always unique, but $\hat{\theta}(w)$ is only
defined modulo $2\pi$. We will thus focus  on $\hat{\theta}'(w)$ which
is unambiguous. This is also the reason why it is preferable to
consider the Burgers equation \eqref{eq:CxBurgers} instead of the
equation \eqref{10} for the potential. 

The effective disorder correlators for the potential and the phase are then defined by
\begin{eqnarray}
\nonumber
\Delta_{V}(w_1-w_2)&:=& \overline{\hat{V}'(w_1)\hat{V}'(w_2)} \\
\label{eq:DefDelta}
\Delta_{\theta}(w_1-w_2)&:=&\overline{\hat{\theta}'(w_1)\hat{\theta}'(w_2)}\ .
\end{eqnarray}
The cross-correlator
$\overline{\hat{V}'(w_1)\hat{\theta}'(w_2)}$ vanishes since $V(x)$ and $\theta(x)$ are independent and due to the symmetry $\theta \to - \theta$. In more general situations, this may not hold. 

The correlators defined above have a nice representation as
correlation functions. Define the normalized expectation value of an observable
$\cal O$ for a given $w$ as 
\begin{eqnarray}\label{}
\left< {\cal O} (x) \right>_{w}&:=& \frac{1}{Z (w)} \sqrt{\frac{\beta m^2}{2\pi}} \\
&& \times 
\int_{-\infty}^{\infty}\rmd x\, e^{-\beta
\big[V(x)+\frac{m^2}{2}(x-w)^{2}\big] -i \theta(x)} {\cal O} (x)\nn \ .
\end{eqnarray}
By definition $\left< 1 \right>_{w}=1$. 
Taking a derivative of \eqref{eq:DefEffPot} yields 
\begin{equation}\label{}
\beta \hat{V}'(w) + i\hat{\theta}'(w) = { \beta m^{2} } \left< w-x \right>_{w} 
\end{equation}
This gives two simple relations for
$\Delta_{V}$ and $\Delta_{\theta}$: 
\begin{eqnarray}
\label{eq:DefCorrZZ}
\Delta_{ZZ}(w_1-w_2) &:=& m^4\overline{\left\langle x-w_1\right\rangle_{w_1} \left\langle x-w_2\right\rangle_{w_2}}  \\
\nonumber &=& \Delta_{V}(w_1-w_2) - \beta^{-2} \Delta_{\theta}(w_1-w_2) \\
\label{eq:DefCorrZZs}
\Delta_{ZZ^*}(w_1-w_2) &: =& m^4\overline{\left\langle x-w_1\right\rangle_{w_1} \left\langle x-w_2\right\rangle_{w_2}^*}  \\
\nonumber &=& \Delta_{V}(w_1-w_2) + \beta^{-2} \Delta_{\theta}(w_1-w_2)
\end{eqnarray}
In terms of the complex Burgers equation \eqref{eq:CxBurgers}, the
effective disorder correlators have  the  intuitive interpretation of equal-time velocity correlation functions:
\begin{eqnarray}
\nonumber
\Delta_{ZZ}(w_1-w_2) &=& \overline{u(w_1,t)u(w_2,t)} \\
\nonumber
\Delta_{ZZ^*}(w_1-w_2) &=& \overline{u(w_1,t)u^*(w_2,t)}
\end{eqnarray}
We  now proceed to compute $\Delta_{ZZ}$ and $\Delta_{ZZ^{*}}$ in each
of the three phases,
and discuss their implications on the physics.

\section{The strong-interference phase (phase III)\label{sec:Ph3}}
The strong-interference phase has  first been discussed in the context
of directed paths with random complex weights in
\cite{Medina1989,Derrida1993,Cook1990} and later for the random-energy
model at complex temperature \cite{Derrida1991a}. In this phase, the
average of $Z$ is essentially zero (or at least subdominant) due to
strong interference, and $Z$ is dominated by fluctuations. The whole
system contributes to the partition sum, in contrast to the case of a real random potential, where it is dominated by a
few points with exceptionally large moduli.

In a replica formalism, this is reflected by a pairing of the replicas, as already observed for the NSS model in \cite{Medina1989}. For the two-dimensional model discussed there, an entropic attraction between replica pairs arises at crossings of four or more replicas due to the spatial structure. In our one-dimensional model, the resulting replica pairs will turn out to be essentially non-interacting and spread out over the whole system.

We will analyze this phase by setting $V(x)=0$ in
\eqref{eq:IntroZ} and consider the small-$m$ limit. We shall show that: (i)
this phase is characterized by $Z(w)$ being 
a Gaussian stochastic process in the complex plane with $w$ as the time
variable; (ii) its two-time correlation function is universal and
given by 
\begin{equation}
\overline{Z(w)Z(w')} \sim e^{-\frac{\beta m^2}{4}(w-w')^2}\ .
\end{equation}
From this, the effective disorder correlators defined  above can be computed. We shall see that $\Delta_V$ and $\Delta_\theta$ exhibit a logarithmic singularity around zero, describing the statistics of zeroes of $Z$. In contrast, $\Delta_{ZZ}=\Delta_V - \beta^{-2}\Delta_\theta$ remains regular around zero.

We then consider two explicit examples where the random phase
disorder is sufficiently strong to observe this phase. Example 1 will
be a model with Brownian imaginary disorder, i.e. long-range correlated phases  $\overline{[\theta(x)-\theta(x')]^{2}}\sim |x-x'|$. Example 2 will be a model with short-range correlated phases uniformly distributed on $[-\pi,\pi[$. 

\subsection{Characterization of phase III and probability distribution of $Z$\label{sec:Ph3ProbDist}}

We set $V(x)=0$ in \eqref{eq:IntroZ} and consider imaginary disorder.
There is a large class of processes $\theta(x)$ such that in the limit $m\rightarrow 0$ the
distribution of $Z(w)$ tends to a complex Gaussian variable due to a central limit theorem (CLT).
To understand qualitatively why, let
us think of $Z(w)$ as a discrete sum $Z(w) \approx \frac{1}{L}  \sum_{j=1}^{L} z_j$
where each $z_j=e^{i \theta_j}$ is a random variable inside the unit disk and $L \sim 1/m$. 
The usual statement of the CLT shows that uncorrelated variables $z_j$ belong to this class,
(this is applied e.g.\ in example 2, section \ref{sec:Ph3Ex2}). In the more general case of correlated
$z_j$ a CLT also holds under the assumption that the correlations of the $z_j$ decay fast enough.
A precise mathematical statement of the necessary and sufficient conditions is possible using a so-called \textit{strong mixing condition} (see \cite{Dehling1986,Bradley2005,Rosenblatt1956}).

More qualitatively, we require the conditions that
\beq \label{qpm}
q_{\pm} = \int_{-\infty}^{\infty} \rmd x\,\overline{e^{i \theta(0) \pm i \theta(x)}}
\eeq
are finite and that a similar condition for the integral of the fourth cumulant holds. 
Since we assumed that $\theta(x)$ is symmetrically distributed around $0$, the 
$q_{\pm}$ are real. 
Note that the fact that $Z(w)$ is bounded for any realization of $\theta(x)$ and any $w$ by
$|Z(w)| \leq 1$ distinguishes this problem from the real potential case, where the
CLT does not hold in general. 

Thus, from now on we consider the case where the CLT holds and in the limit $m\rightarrow 0$ the
distribution of $Z(w)$ tends to a complex Gaussian variable. This happens
in what we call phases I and III. In these phases, the distribution of
$Z(w)$ is thus determined by its mean
$\overline{Z(w)}$ and its covariance matrix, consisting of 3 entries
$\overline{Z(w)Z(w)}$, $\overline{Z(w)Z^*(w)}$, and $\overline{Z^*(w)Z^*(w)}$.
A similar reasoning applies to the joint distribution of $Z(w)$ and $Z(w')$.

The key difference between the strong-interference phase III and the
high-temperature phase I is  the scaling of these
moments: If the mean of $Z$ as a function of $m$ decreases faster than the
fluctuations, $\overline{Z(w)}^2 \ll
\overline{(Z(w)-\overline{Z(w)})^2} \sim m$ as $m\rightarrow 0$, we
obtain the strong-interference, fluctuation-dominated phase III. If,
on the other hand, the mean decreases slower than the fluctuations, $\overline{Z(w)}^2 \gg m$ as $m\rightarrow 0$, we obtain the high-temperature phase I.

In the example in section \ref{sec:Ph3Ex1}, we will take $\theta$ to
be long-range correlated, $\overline{[\theta(x)-\theta(y)]^2}\sim |x-y|$, whence we
will see that $\overline{Z(w)}^2 \sim e^{-\alpha / m} \ll m$. On the
other hand, in section \ref{sec:Ph3Ex2}, we will consider an example where rotational symmetry  enforces $\overline{Z(w)}=0$. In both cases, we verify the general results and assumptions presented here.

\subsection{The second moments \label{sec:Ph3SecondMoment}}

The second moments of the complex process $Z(w)$ take a general
form which we derive now. The ``renormalization group'' equation (\ref{9}) describes how $Z (w)$
evolves under changes of $t=m^{-2}$. This implies a similar equation
for the $2$-point function $\tilde{f}(w-w'):=\overline{Z(w)Z^*(w')}$,
\beq
\label{eq:FlowTildeF}
\partial_{t} \tilde{f}_t(w) = T\partial_{w}^{2}\tilde{f}_t(w)
\eeq
Here, we added the index $t$ to make the dependence of $Z(w)$ on the parameter $m$ and hence the dependence of $\tilde{f}(w)$ on the parameter $t$ explicit. The general solution of \eqref{eq:FlowTildeF} in terms of the initial condition $\tilde{f}_0(w)=\overline{e^{i (\theta(0) - \theta(w))}}$ is
\begin{equation}
\label{eq:SolFlowTildeF}
\tilde{f}_t(w) = \sqrt{\frac{1}{4\pi T t}}\int_{-\infty}^{\infty}
e^{-\frac{(w-w_0)^2}{4 T t}}\tilde{f}_0 (w_0) \, \rmd w_0\ .
\end{equation} 
Since we assumed $q_-$ to be finite, see Eq. (\ref{qpm}), the solution \eqref{eq:SolFlowTildeF} tends to a Gaussian scaling form as $t\rightarrow\infty$
\beq
\label{eq:TildeFAsympt}
\tilde{f}_t(w) \rightarrow q_{-}\sqrt{\frac{1}{4\pi T t}} e^{-\frac{w^{2}}{4 T t} }\ .
\eeq
Note that this assumption is violated in phase I, where the mean $\overline{Z(w)}$ contributes a constant to $f_t(w)$ even for large $t$. It is also violated in phase II, where $\int_{-\infty}^\infty \tilde{f}_0(w_0) dw_0$ diverges.

Going back to the original variables $m$ and $\beta$, the asymptotic
form of \eqref{eq:SolFlowTildeF} in phase III is 
\beqn
\label{eq:FlowScaling}
\tilde{f}_m(w-w') & \stackrel{m\rightarrow 0}{-\!\!\!\longrightarrow} & q_-\sqrt{\frac{\beta}{2\pi}}m f\big(\hat{w}=m\sqrt{\beta} (w-w')\big)\ \ \ \ \ \\ 
\label{eq:FlowFPF}
f(\hat{w}) &=& e^{-\frac{1}{4}\hat{w}^2}\ .
\eeqn
Exactly the same reasoning goes through for the second moment $\overline{Z(w)Z(w')}$ with $q_-$ replaced
by $q_+$.

The scaling in \eqref{eq:FlowScaling} reflects the fluctuation-driven
character of phase III: If the mean $\overline{Z(w)}$ were not subdominant, for large
system sizes $L\sim m^{-1}$, as
compared to the
fluctuations, $\tilde{f}$ in \eqref{eq:FlowScaling} would be
$\mathcal{O}(L^0)$ instead of $\mathcal{O}(L^{-1})$, and not tend to
zero for large argument.

To summarize, in the strong-interference phase III, as $m\rightarrow
0$, the partition function $Z(w)$ tends to a Gaussian  process with
mean 0 and correlation matrix: 
\begin{widetext}
\begin{align}
\label{eq:Ph3SecondMoment}
& \left( 
\begin{array}{cccc}
	\overline{Z(w)Z(w)} & \overline{Z(w)Z^*(w)} & \overline{Z(w)Z(w')} & \overline{Z(w)Z^*(w')} \\
	\overline{Z^*(w)Z(w)} & \overline{Z^*(w)Z^*(w)} & \overline{Z(w)Z(w')} & \overline{Z(w)Z^*(w')} \\
	\overline{Z(w')Z(w)} & \overline{Z(w')Z^*(w)} & \overline{Z(w')Z(w')} & \overline{Z(w')Z^*(w')} \\
	\overline{Z^*(w')Z(w)} & \overline{Z^*(w')Z^*(w)} & \overline{Z^*(w')Z(w')} & \overline{Z^*(w')Z^*(w')}
\end{array}
\right)   = m\sqrt{\frac{\beta}{2\pi}}\left( 
\begin{array}{cccc}
	q_+  & q_-  & q_+ f(\hat{w}) & q_- f(\hat{w}) \\
	q_-  & q_+  & q_- f(\hat{w}) & q_+ f(\hat{w}) \\
	q_+ f(\hat{w}) & q_- f(\hat{w}) & q_+ & q_-  \\
	q_- f(\hat{w}) & q_+ f(\hat{w}) & q_- & q_+
\end{array}
\right)
\end{align}\end{widetext} 

With this, we have completely characterized the $m\rightarrow 0$ asymptotics of $Z(w)$ in the strong-interference phase III as a Gaussian stochastic process with the second moment given by \eqref{eq:FlowScaling}.
In sections \ref{sec:Ph3Ex1} and \ref{sec:Ph3Ex2} we shall explicitely check the asymptotic form in \eqref{eq:FlowFPF} and obtain the non-universal constant $q_{\pm}$ in \eqref{eq:FlowFPF}.

\subsection{The disorder correlators}
Having discussed the probability distribution of $Z(w)$, we now turn
to computing the effective disorder correlators. For simplicity, we  restrict
ourselves to the case when the limiting Gaussian distribution for
$Z(w)$ is rotationally symmetric, i.e.\ only depends on the modulus
$|Z(w)|$. In the covariance matrix \eqref{eq:Ph3SecondMoment}, this
means $q_+=0$. The joint probability distribution for two partition
sums $Z(w_1)=a_1+i\, b_1$, and $Z(w_2)=a_2+i\, b_2$ is then given by
\begin{equation}
\label{eq:GaussianDist4d}
P(a_1,b_1,a_2,b_2)=\frac{1}{4\pi^2 \sqrt{\det{B}}}e^{-\frac{1}{2}\vec{x}B^{-1}\vec{x}}\ ,
\end{equation}
with $\vec{x}= \left(\begin{array}{cccc}
	a_1 & b_1 & a_2 & b_2 \end{array} \right)$, 
\begin{equation}
\label{eq:GaussianCorr4d2}
B= \frac{c}{2} \left(\begin{array}{cccc}
	1 & 0 & f(\hat{w}) & 0 \\
	0 & 1 & 0 & f(\hat{w}) \\
	f(\hat{w}) & 0 & 1 & 0 \\
	0 & f(\hat{w}) & 0 & 1
\end{array}\right)\ ,
\end{equation} 
and $c=mq_-\sqrt{\frac{\beta}{2\pi}}$. We will see later that the disorder correlator does not depend on $c$.

To compute the effective disorder correlators, let us reconsider their
definition  \eqref{eq:DefDelta}. Since $\hat{\theta}$ is the angular
variable of a two-dimensional Gaussian stochastic process, we can
apply the results of \cite{LeDoussal2009}. There, the two-time
correlation function for the angular ``velocity'' $\dot{\hat \theta} (w):=
\partial_{w} \hat \theta (w)$  of planar Brownian motion is (cf.\
\cite{LeDoussal2009a}, formula 17):
\begin{equation}\label{eq:CorrTT}
{\overline{\dot{\hat \theta}(w)\dot{\hat \theta}(w')}} =-\frac{1}{2}\left[\partial_w \partial_{w'}\log{f(\hat{w})}\right] \log{\left(1-f(\hat{w})^2\right)}  \ .
\end{equation}
The two-point correlator of the  phase  (instead of its velocity) can then be written as a double integral of \eqref{eq:CorrTT}, but no explicit expression is known.

The two-point correlator of $\ln |Z|=\beta \hat{V} $ is obtained
from the explicit form \eqref{eq:GaussianDist4d} for the two-time
probability distribution as 
\begin{eqnarray}
 \nonumber\lefteqn{\beta^{2}\overline{\hat{V} (w)\hat{V} (w')}  =  \overline{\log{|Z(w)|}\log{|Z(w')|}} } & & \\
\nonumber &=& \int_{-\infty}^\infty \frac{1}{4\pi^2 \sqrt{\det{B}}}e^{-\frac{1}{2}\vec{x}B^{-1}\vec{x}} \log{|a_1 + i\,b_1|} \log{|a_2 + i\,b_2|}
\end{eqnarray}
with $B$ given by \eqref{eq:GaussianCorr4d2}. 
This integral can be computed exactly ($\gamma_{\mathrm{E}}$ denotes
Euler's constant):
\begin{equation}
\label{eq:CorrRR}
\beta^{2}\overline{\hat{V} (w)\hat{V} (w')} = \frac{1}{4}\left[\left(\gamma_{\mathrm{E}}-\log{c}\right)^2 + \text{Li}_2 (f(\hat{w} )^2)\right]
\end{equation}
Plugging in the scaling form $f(\hat{w})=e^{-\frac{1}{4}\hat{w}^2}$
into \eqref{eq:CorrTT} and \eqref{eq:CorrRR}, we obtain the disorder
correlators 
\begin{eqnarray}
\nonumber
\Delta_{V}(w) &=& -\frac{m^2}{4\beta}\left[\frac{ \hat{w}^2}{e^{\frac{1}{2}\hat{w}^2}-1} + \log{\left(1-e^{-\frac{1}{2}\hat{w}^2}\right)}\right] \\
\label{eq:GaussianDeltaWW}
\Delta_{\theta}(w) &=& -\beta\frac{m^2}{4} \log{\left(1-e^{-\frac{1}{2}\hat{w}^2}\right)}\ .
\end{eqnarray}
Equivalently, 
\begin{eqnarray}
\label{eq:GaussianDeltaZZ}
\Delta_{ZZ}(w) &=& -\frac{m^2}{4\beta}\frac{ \hat{w}^2}{e^{\frac{1}{2}\hat{w}^2}-1} \\
\label{eq:GaussianDeltaZZs}
\Delta_{ZZ^*}(w) &=& -\frac{m^2}{4\beta}\left[\frac{ \hat{w}^2}{e^{\frac{1}{2}\hat{w}^2}-1} + 2 \log{\left(1-e^{-\frac{1}{2}\hat{w}^2}\right)}\right]\ .\qquad 
\end{eqnarray}
Observe that $\Delta_{ZZ}$ is smooth around $0$, whereas $\Delta_{ZZ^*}$ has a logarithmic singularity at ${w}=0$.
Note that all correlators are expressed in terms of the rescaled variable $\hat w$ defined in \eqref{eq:FlowScaling}. 

The above expressions for the correlators (which are also the two-point equal-time velocity correlators for the complex
Burgers equation) are universal and generally valid in phase III, under the assumption of rotationally invariant disorder.  The more general case can be handled by similar methods but is not studied here. These results were obtained
using the CLT assumption.

We now study two specific models where we can compute the general moments (beyond the second one) using the replica method, and check that they are consistent with the above reasoning. As an additional check we also compute numerically the correlators.

\subsection{Example 1: Imaginary Brownian disorder \label{sec:Ph3Ex1}}
\begin{figure}
	\centering
		\includegraphics{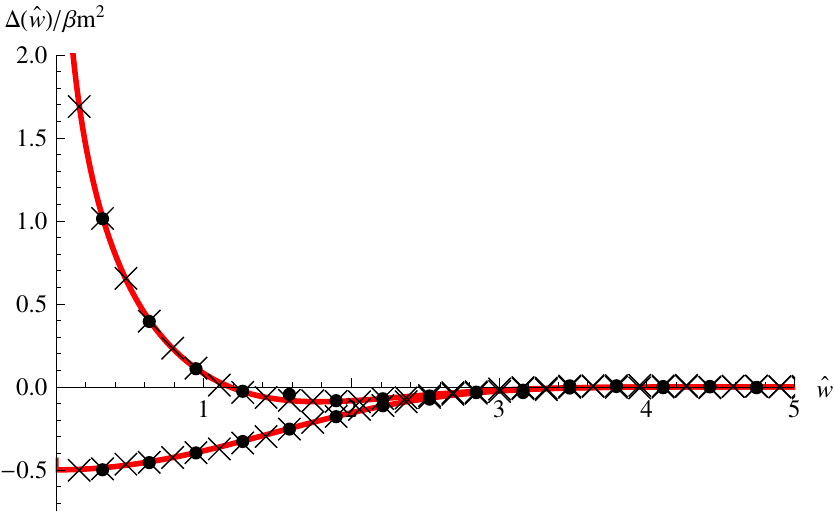}
	\caption{Effective force-force correlators for the long-range model defined by \eqref{eq:LRWDef}. Red line: $\Delta_{ZZ}$ from \eqref{eq:GaussianDeltaZZ}, blue line: $\Delta_{ZZ^*}$ from \eqref{eq:GaussianDeltaZZs}. Dots: correlators obtained from numerical simulations using \eqref{eq:DefCorrZZ} and \eqref{eq:DefCorrZZs} for $\sigma=1, \beta=10$ and $m=0.05$ (blue and purple), $m=0.1$ (yellow and green).}
	\label{fig:CorrLR}
\end{figure}

Consider pure random-phase disorder, $V(x)=0$, and take $\theta(x)$ to be a continuous random walk, i.e.\ a Gaussian stochastic process satisfying
\begin{equation}
\label{eq:LRWDef}
\overline{[\theta(x)-\theta(x')]^2} = 2 \sigma |x-x'|\ .
\end{equation}
Thus in the finite length regularization, the partition sum \eqref{eq:CxPartSumL} is the Sinai model at an imaginary temperature.
We measure numerically the effective disorder correlators, 
using  relations \eqref{eq:DefCorrZZ} and \eqref{eq:DefCorrZZs}.
The results are compared in figure \ref{fig:CorrLR} against the
analytic computation in the previous section. We observe good agreement.


\subsubsection{Second moment}
Here we show explicitely the validity of the scaling argument given in section \ref{sec:Ph3SecondMoment} for this model.

Using  formula \eqref{eq:IntroZ}, the second moment is given by
\begin{eqnarray}\label{eq:CxSecondMoment1}
\lefteqn{\overline{Z (w) Z^*(w')}}  \\
&&= \frac{\beta m^2}{2\pi}\int_{-\infty}^\infty \rmd x\,\rmd y \, e^{-\frac{1}{2}\sigma|x-y| -\beta\frac{m^2}{2}\left[(x-w)^2+(y-w')^2\right]}\nonumber 
\end{eqnarray}
This integral \eqref{eq:CxSecondMoment1} can be computed exactly by
using the center-of-mass variable $s$ and the ``pair separation''
variable $t$, defined as
\begin{eqnarray}
\nonumber s &:=& \frac{x+y}{2}\\
\label{eq:SplitCoord} t &:=& x-y\ .
\end{eqnarray}
In these variables, the $s$ and $t$-integrals decouple,
\begin{eqnarray}
\nonumber
\overline{Z (w) Z^*(w')} &=& \frac{\beta m^2}{2\pi} \left(\int_{-\infty}^\infty \rmd s\,e^{-\beta \frac{m^2}{2} \left[(s-w)^2+(s-w')^2\right]}\right) \\
\label{eq:CxSecondMoment2}
& & \times \left(\int_{-\infty}^\infty \rmd
t\,e^{-\frac{1}{2}\sigma|t|+ \beta\frac{m^2}{2}(w-w')t-\beta
\frac{m^2}{4} t^2}\right) \nn \\
\end{eqnarray}
Both integrals can  be performed analytically. In terms of the
rescaled variables  
\begin{eqnarray}
\label{eq:DefWHat}\hat{w} &:=& m \sqrt{\beta} (w-w') \\
\nonumber\hat{\sigma} &:=& \frac{\sigma}{m\sqrt{\beta}},
\end{eqnarray}
the second moment \eqref{eq:CxSecondMoment2} is  given by
\begin{eqnarray}
\nonumber
\overline{Z (w) Z^*(w')} = &\frac{1}{2} e^{-\frac{\hat{w}^2}{4}} &\left[ e^{\frac{(\hat{\sigma}-\hat{w})^2}{4}} \text{Erfc}\left( \frac{\hat{\sigma}-\hat{w}}{2} \right)\right. \\
\label{eq:CxSecondMoment3}
& & \left. +e^{\frac{(\hat{\sigma}+\hat{w})^2}{4}} \text{Erfc}\left( \frac{\hat{\sigma}+\hat{w}}{2} \right) \right]
\end{eqnarray}
Taking the $m\rightarrow 0$ limit at fixed $\beta$ (i.e.\ the limit $\hat{\sigma}\rightarrow \infty$), the second moment \eqref{eq:CxSecondMoment3}  approaches the scaling form
\begin{eqnarray}
\nonumber
\overline{Z (w) Z^*(w')} &\sim& \frac{2}{\sqrt{\pi}\hat{\sigma}} f(\hat{w})\\
\label{eq:CxSecondMoment4}
f(\hat{w}) &=& e^{-\frac{1}{4}\hat{w}^2}
\end{eqnarray}
This is exactly the scaling form obtained in \eqref{eq:FlowFPF}, and gives a non-trivial check for the validity of that argument.

\subsubsection{Higher moments}
Let us now look at higher moments of $Z$ and $Z^*$, given by:
\begin{eqnarray}\label{eq:CxHigherMoments}
\lefteqn{\overline{Z (w_1)...Z(w_n) Z^*(w'_1)...Z^*(w'_n)}}\nn \\
&& = \int\limits_{-\infty}^\infty \rmd x_1\dotsb  \rmd x_n
\int\limits_{-\infty}^\infty \rmd y_1\dotsb  \rmd y_n\,\nn \\
&&\qquad  \times e^{-\frac{\sigma}{4}\left[\sum_{i,j=1}^n{|x_i-y_j| + |y_i-x_j| -
|x_i-x_j| - |y_i-y_j|}\right]} \nn \\
&&\qquad \times e^{ -\beta\frac{m^2}{2}\sum_{i=1}^n\left[(x_i-w_i)^2+(y_i-w'_i)^2\right]}
\end{eqnarray}
An exact calculation does not seem feasible, but the asymptotic
behaviour is understood as follows: 
In the limit $\hat{\sigma} \rightarrow \infty$, the exponent in
\eqref{eq:CxHigherMoments} will have a sharp maximum at configurations
where the $x_{i}$ and $y_{j}$ are paired, i.e. close to each others. We now consider configurations which are close to such a pairing, where
w.l.o.g.\  $x_i$ is paired to $y_{\pi(i)}$ with some permutation $\pi$. Similar to \eqref{eq:SplitCoord} we introduce center-of-mass and separation coordinates $s_i$ and $t_i$ for each pair  and rewrite the mass terms as in \eqref{eq:CxSecondMoment2}.

The $t_i$-integrals  have  complicated boundaries, which  
yield terms decaying as $e^{-\alpha \hat{\sigma}}$ with various
functions $ \alpha > 0$. Hence, these terms can be neglected in the
limit $m\rightarrow 0$, and the $s$ and $t$-integrals decouple again:
\begin{eqnarray}\label{eq:LRHigherMomentsGen}
\lefteqn{\overline{Z (w_1)...Z(w_n) Z^*(w'_1)...Z^*(w'_n)}}\nn \\
&=&  \sum_{\pi} \prod_{i=1}^n \overline{Z(w_i)Z^*(w'_{\pi(i)})}+\text{ higher orders in }  m\qquad 
\end{eqnarray}
In particular, we get
\begin{eqnarray}\label{eq:LRHigherMoments}
\overline{\left[Z (w) Z^*(w')\right]^n}& =& n! \left[ \overline{Z (w)Z^*(w')} \right]^n \nn \\
&&+ \text{ higher orders in }  m\qquad 
\end{eqnarray}
A more rigorous justification that this is the leading term in an expansion in orders of $m$ is given in appendix \ref{sec:AppLMoments} by considering the moments of the partition sum in a finite system \eqref{eq:CxPartSumL}.

Correspondingly, the leading term for the moments $\overline{\left[Z
(w)\right]^n \left[Z^*(w')\right]^m}$ for $m\neq n$ is zero in the
strong-disorder limit, since then the replicas cannot be
paired. Stated differently,  the phase of $Z$ is random, and hence only moments invariant under the rotation $Z\rightarrow e^{i\phi}Z$ are nonzero.
Dropping the higher-order terms in \eqref{eq:LRHigherMomentsGen}, we obtain exactly the moments of a complex Gaussian variable. This supports
the general claim made in section \ref{sec:Ph3ProbDist}, and shows
that this model is indeed in the strong-interference phase III.

The fact that configurations with unpaired replicas are subdominant
shows that fluctuations of $Z$ dominate over the average. 
Intuitively, this happens since for $\hat{\sigma} \gg 1$ the phase of
the integrand in the expression \eqref{eq:IntroZ} grows beyond $2\pi$
on a scale much smaller than the width $\frac{1}{m}$ of the parabolic well. Hence, \eqref{eq:IntroZ} is essentially a sum of many random complex numbers with mean $ 0$.

This is  the same behaviour as in  ``phase III'' discussed by Cook
and Derrida  \cite{Cook1990} and by Derrida
\cite{Derrida1991a}. Since our potential is   long-range correlated,
$\overline{\theta(x)^2}\sim x$ instead of the short-range correlated
potential $\overline{\theta(x)^2}\sim 1$ used in \cite{Cook1990}, the
complex phase of the integrand in \eqref{eq:IntroZ} grows much faster
in our model. Hence, we do not observe ``phase I'' for high
temperatures ($\beta < \beta_c$) as in \cite{Cook1990,Derrida1991a}, but only the fluctuation-dominated ``phase III''.

In the following, we shall show that similar results hold in a model
with uniformly distributed $\theta (x)$. 

\subsection{Example 2: A short-range correlated model with uniformly distributed angles\label{sec:Ph3Ex2}}
Our second example  is a model, where the potential $\theta(x)$ in \eqref{eq:CxPartSumL} is short-range correlated and uniformly distributed. To be more precise,
\begin{eqnarray}
\label{eq:SRWDef}
P\big(\theta(x)\big) = \frac{1}{2\pi}\ ,
\end{eqnarray}
while $\theta(x)$ and $\theta(x')$ are uncorrelated for $x \neq x'$.

\begin{figure}
	\centering
		\includegraphics{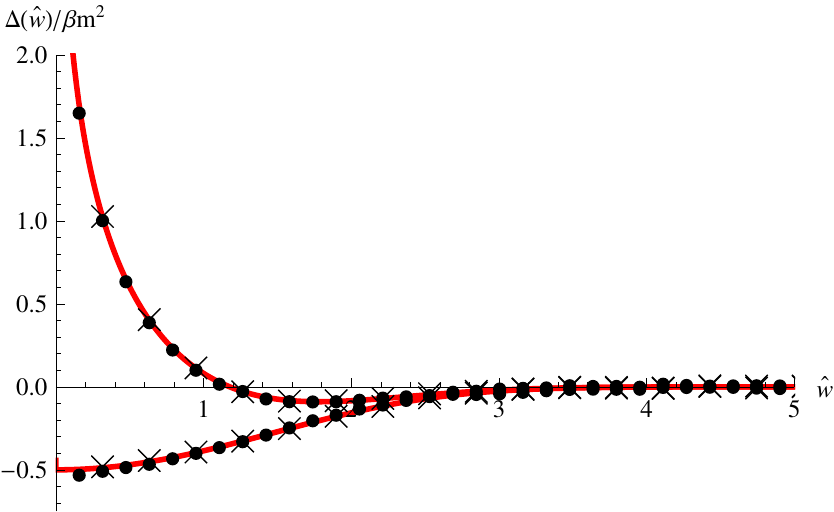}
	\caption{Effective force-force correlators for the short-range model defined by \eqref{eq:SRWDef}. Red line: $\Delta_{ZZ}$ from \eqref{eq:GaussianDeltaZZ}, blue line: $\Delta_{ZZ^*}$ from \eqref{eq:GaussianDeltaZZs}. Dots: correlators obtained from numerical simulations using \eqref{eq:DefCorrZZ} and \eqref{eq:DefCorrZZs} for $\sigma=1, \beta=10$ and $m=0.05$ (blue and purple), $m=0.1$ (yellow and green).}
	\label{fig:CorrSR}
\end{figure}

As can be seen on figure \ref{fig:CorrSR}, a numerical simulation
yields disorder correlators which compare  well to the general results
obtained above. As for the first example, we shall compute  moments of $Z$ to elucidate the physics.

Invariance of the distribution of $\theta (x)$ under a phase shift, $\theta(x)\rightarrow\theta(x)+\phi$, implies invariance of the distribution of $Z$ under $Z\rightarrow Z\,e^{i\phi}$. Hence, the only nonzero moments are of the form $\overline{|Z(w)|^{2n}}$.

For $n=1$, evaluating the second moment gives:
\begin{eqnarray}
\nonumber
\lefteqn{\overline{Z (w) Z^*(w')} =} & & \\
\nonumber
&=& \frac{\beta m^2}{2\pi}\int_{-\infty}^{\infty} e^{-\beta \frac{m^2}{2}\left[(x-w)^2+(x-w')^2\right]} \overline{e^{-i[\theta  (x)-\theta (x')]}}\,\rmd x\, \rmd x' \\
\nonumber
&=& \frac{\beta m^2}{2\pi}\int_{-\infty}^{\infty} e^{-\beta \frac{m^2}{2}\left[(x-w)^2+(x-w')^2\right]}\,\rmd x \\
\label{eq:SRSecondMoment1}
&=& \sqrt{\frac{\beta m^2}{\pi}}\frac{1}{2}e^{-\beta\frac{m^2}{4}(w-w')^2}
\end{eqnarray}
Note that this is again in agreement with the general scaling
argument given in section \ref{sec:Ph3SecondMoment}.

For the higher moments $\overline{\left[Z (w) Z^*(w')\right]^n}$,  the
only terms contributing  are those where the $2n$ replica form $n$
pairs. When at least two pairs are at the same position, it is not
clear which replicas are pairs, leading to double-counting. However
these contributions are subdominant and vanish with a relative factor
of at least $m^{2}$. Thus, the dominant term for $m\rightarrow 0$ is
\begin{eqnarray}
\nonumber
\lefteqn{\overline{\left[Z (w)Z^*(w')\right]^n}}& & \\
\nonumber
&=&n!\sqrt{\frac{\beta m^2}{2\pi}}^{2n} \left[\int_{-\infty}^\infty e^{-\beta \frac{m^2}{2}\left[(x-w)^2+(x-w')^2\right]}\rmd x\right]^n\\
\nonumber
&=& n! \left[\overline{Z (w) Z^*(w')}\right]^n + \text{higher orders in } m
\end{eqnarray}
Analogously to the derivation of \eqref{eq:LRHigherMoments} this
formulas can be generalized to moments of $Z$ with different
positions. 

Again, we observe the same behaviour of the moments as for a complex Gaussian. In total, in the limit of large $m$, we recover the same phase III results as in the long-range correlated model in section \ref{sec:Ph3Ex1}, and confirm the validity of the general arguments in the beginning of this section once more.

\section{The frozen phase (phase II)\label{sec:Ph2}}
For large $\beta$ and sufficiently strong potential $V(x)$, the
modulus of the integrand in \eqref{eq:IntroZ} has a very broad
distribution. It is well-known that the partition sum (in the absence
of the harmonic well) is then dominated by a few points, the minima of
$V(x)$. This so-called \textit{frozen} phase has been extensively
studied in the absence of random phases by a variety of methods
(replica symmetry breaking \cite{Bouchaud1997}, functional
renormalization group \cite{LeDoussal2009} and rigorous mathematical analysis \cite{Derrida1981})

Distributions of $V$ where a frozen phase occurs in the model \eqref{eq:IntroZ} in absence of complex phases include:
\begin{itemize}
	\item Long-range correlated random potentials $V(x)$, i.e.\
	$\overline{V(x)V(x')}=\sigma |x-x'|$. This is known as the  \textit{Sinai model}, which describes the diffusion of a random walker in a 1D random static force field \cite{Sinai1982,LeDoussal2003}.
	\item Short-range correlated random potentials $V(x)$, i.e.\
	$\overline{V(x)V(x')}=\sigma \delta(x-x')$, where the
	amplitude is rescaled logarithmically with the system size, or
	$m$: $\sigma \sim -\log m$. Freezing occurs  below some
	critical temperature, $\beta > \beta_c$, analogously to  the random energy model \cite{Derrida1981}.
\end{itemize}
Among the most interesting features of the frozen phase is the appearance of jumps between distant minima of $V(x)$ as the position $w$ of the harmonic well in \eqref{eq:IntroZ} is varied \cite{LeDoussal2006,LeDoussalWiese2006a,LeDoussalMiddletonWiese2008,LeDoussalWiese2008c}. These correspond to shocks \cite{Bec2007} under the mapping to the Burgers equation discussed in section \ref{sec:CxBurgers}. 
In the following, we will discuss how their structure is changed upon introduction of random complex phases $\theta(x)$, following the standard treatment \cite{Bernard1998,LeDoussal2009} for the case without random phases.

\subsection{Complex shocks\label{sec:Ph2Shocks}}

Let us first consider a fixed realization of the random potential $V(x)$
and the random phases $\theta(x)$. For almost all $w$, the real part of
the exponent, $V(x)+\frac{m^2}{2}(x-w)^2$ has, as a function of $x$,  a single minimum at some value $x=x_m(w)$. Then, in the low-temperature limit (i.e.\ $\beta \rightarrow \infty$)
\begin{equation}
Z(w) = e^{-\beta V(x_m) - i\theta(x_m) - \beta\frac{m^2}{2}(x_m-w)^2}
\end{equation}
and hence 
\begin{eqnarray}
\label{eq:FrozenVhat}
\hat{V}(w) &=& V(x_m) + \frac{m^2}{2}(x_m-w)^2 \\
\hat{\theta}(w) &=& \theta(x_m)
\end{eqnarray}
The function $x_{m} (w)$ is constant over some range of $w$, but then
jumps to a different value at $w=w^{*}$. Denoting the two solutions at
$w^{*}$ by  $x_1$ and  $ x_2$, the necessary condition for a jump is 
\begin{equation}\label{47}
V(x_1)+\frac{m^2}{2}(x_1-w^*)^2= V(x_2)+\frac{m^2}{2}(x_2-w^*)^2
\end{equation}
In terms of the effective potential $\hat{V}$, two parabolic sections given by \eqref{eq:FrozenVhat} (with $w=w_1$ and $w_2$, respectively) meet at $w^*$ with a linear cusp. The first derivative, $\hat{V}'(w)$, has a discontinuity at $w^*$.

So far, this is  the same picture as has been established for purely real disorder long ago 
in the context of the Burgers equation  \cite{Burgers1974,Bec2007,BalentsBouchaudMezard1996}. There, the appearance of the shocks is succintly encoded \cite{LeDoussal2006,LeDoussal2009}  in
the effective force-force correlator $\Delta(w)$, which extends to the broader context of interfaces in random media.
It has been computed
and tested both numerically
\cite{MiddletonLeDoussalWiese2006,Rosso2007} and experimentally
\cite{LeDoussal2009d}. It encodes the statistics of the shocks through a linear
cusp at $w=0$. At finite temperature $\beta$, the shock is smoothened in the
so-called thermal boundary layer which extends on
a scale $w\sim T=\beta^{-1}$
\cite{Balents2005,ChauveGiamarchiLeDoussal2000}.

The additional random phase $\theta (x)$ will in general be different
at $x_1$ and $x_2$. We now show that this is reflected in the profile
of $\hat{V}(w)$ and $\hat{\theta}(w)$ for $w$ close to a shock. This modifies the form of the 
disorder correlator $\Delta(w)$ near $w=0$, more specifically in the thermal boundary layer region $w \sim T$
where we will obtain its precise form. We find that it adds a logarithmic singularity which depends 
on the statistics of the phase jumps.

\subsubsection{The shock profile -- general case}
\begin{figure*}
\subfigure[]{
   \includegraphics[scale =1] {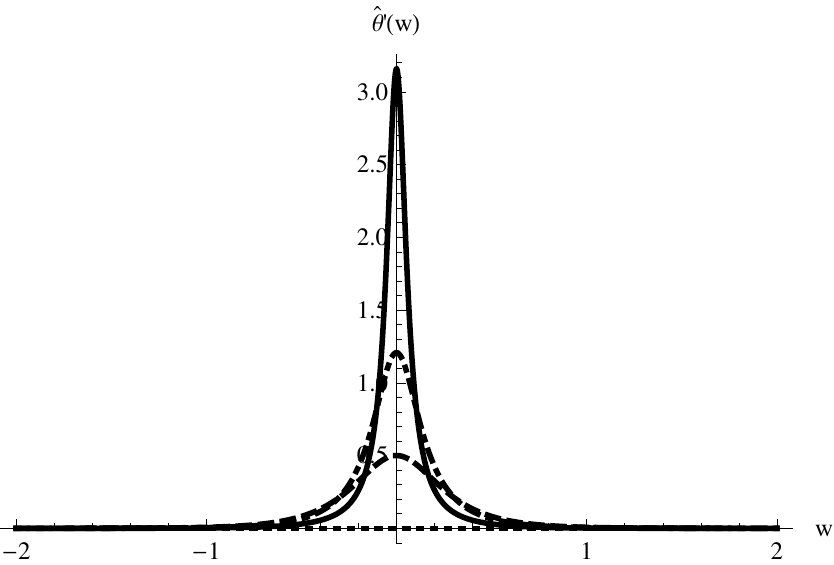}
 }
 \subfigure[]{
   \includegraphics[scale =1] {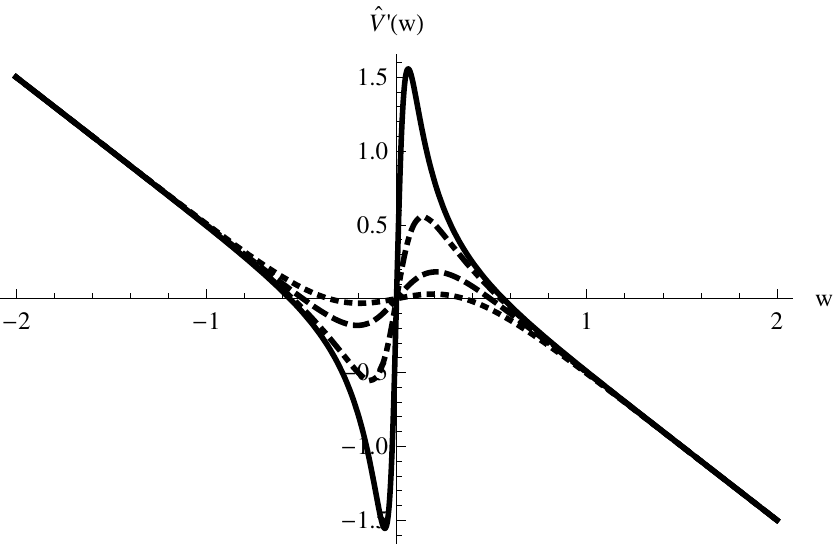}
 }
\label{fig:ShockProfiles}
\caption{Shock profiles from \eqref{eq:Ph2EffPot} for $\beta=5$, $w^*=0$, $x_1+x_2=0$, $m=1$ and $\phi=0$ (dotted), $\phi=\frac{\pi}{2}$ (dashed), $\phi=\frac{3}{4}\pi$ (dot-dashed), $\phi=\frac{9}{10}\pi$ (solid).}
\end{figure*}

Let us assume a two-well picture, i.e.\ approximate $Z(w)$ by
\begin{equation}
Z(w) = e^{-\beta[V_1+\frac{m^2}{2}(x_1-w)^2] - i \theta_1} + e^{-\beta[V_2+\frac{m^2}{2}(x_2-w)^2] - i \theta_2}
\end{equation}
The effective potential \eqref{eq:DefEffPot} can be written in terms of the jump
size $s:=\beta m^2 (x_2-x_1)$, the  phase difference, $\phi: =
\theta_2 - \theta_1$, and $w^{*}$, solution of (\ref{47}):
\begin{eqnarray}
\hat{\theta}'(w) &=& \frac{s}{2} \frac{\sin(\phi)}{\cos(\phi) + \cosh\left ( s [w-w^*]\right)} \\
\label{eq:Ph2EffPot}
 -\hat{V}'(w) &= & \frac{s}{2\beta}\frac{\sinh\left ( s [w-w^*] \right)}{\cos(\phi) + \cosh\left( s [w-w^*] \right)} \\
\nonumber & & + \frac{m^2}{2}(x_1+x_2-2w)
\end{eqnarray}
Some examples of shock profiles for various values of the parameters are shown in figure \ref{fig:ShockProfiles}.
Note that as $\phi \rightarrow \pm \pi$, a pole appears in $\hat{V}'(w)$ at $w=w^*$ which is
the real part of the Burgers velocity.

To obtain the disorder correlator $\Delta_{\theta}$, we need to
average $\hat{\theta}'(w_1)\hat{\theta}'(w_2)$ over  the
disorder. Assume a small uniform density $\rho_0$ of  shocks, and
average over $w^*$ with the measure $\rho_0
\int_{-\frac{1}{2\rho_0}}^{\frac{1}{2\rho_0}} \rmd w^*$. Since
$\hat{\theta}'(w)$ decays  rapidly as $w^*$ is increased, we can
safely extend the integration limits to $\pm\infty$, allowing to
compute the integral over the shock position $w^*$ analytically:
\begin{eqnarray}
\label{eq:CxShockGen}
\Delta_{\theta}(w) &=&  \rho_0 \int_{-\pi}^\pi \rmd \phi \int_{0}^{\infty} \rmd s\,P(\phi, s) f(\phi, s)\qquad  \\
f(\phi, s) & = & s \sin^2(\phi) \frac{\phi \cot \phi - \frac{sw}{2}
\coth \frac{sw}{2}}{\cos 2\phi - \cosh sw}
\end{eqnarray}
where $w=w_1-w_2$. We denote by $P(\phi,s)$ the joint distribution of the jump sizes $s$ and the phase jumps $\phi$.
Remarkably, $\Delta_{V} (w)$ can also be calculated, by considering the difference  $\Delta_{V} (w)-
\Delta_{V,\phi =0} (w)$, where  $\Delta_{V,\phi =0} (w)$ is the
correlator of the problem without the imaginary disorder, $\theta (x)
=0$: 
\begin{equation}\label{}
\Delta_{V}(w)  =  \Delta_{V,\phi=0}(w)  + \beta^{-2} \Delta_{\theta} (w)\ .
\end{equation} 
Thus the correlator $\Delta_{ZZ}=\Delta_V - \beta^{-2}\Delta_\theta$ is unchanged by the complex phases. 
 Observe that the integrand in \eqref{eq:CxShockGen} becomes singular for $w=0$ and $\phi=\pm \pi$. In the following examples (sections \ref{sec:Ph2Ex1} and \ref{sec:Ph2Ex2}) we shall see that this singularity yields a logarithmic singularity in $\Delta_{V,\theta}$ around zero. Its coefficient will be seen in section \ref{sec:Ph2Ex2} to be proportional to $P(\phi=\pm\pi)$.

We have now discussed the effective disorder correlators $\Delta_\theta$ and $\Delta_V$ in a two-well approximation in a general situation. So far, we did not make specific assumptions on the distribution and the correlations of the disorder. These enter the final result \eqref{eq:CxShockGen} through the joint distribution of the jump sizes $s$ and the phase jumps $\phi$.
Now, we will specialize to examples of particular interest.

\subsection{Example 1: Uniformly distributed random phases in a short-range potential\label{sec:Ph2Ex1}}
Our first example is $\theta(x)$ uniformly distributed in $[-\pi,\pi]$
and uncorrelated from the spatial dependence $x_{i}$, i.e.\ $P(\phi, s) = \frac{1}{2\pi} P(s)$. This allows to perform the $\phi$ integral in \eqref{eq:CxShockGen} analytically:
\begin{equation}
\label{eq:DeltaPh2Unif}
\Delta_{\theta}(w) =\rho_0 \int_{0}^{\infty} \rmd s\, P(s) \frac{s}{2}\left[\frac{sw}{e^{sw}-1}-\log\left(1-e^{-sw}\right)\right]
\end{equation}
To take the limit $\beta\rightarrow \infty$, we  write $s=\beta m^2 (x_2-x_1)=\beta m \mu
\hat{s}$, where $\mu$ is the jump-size scale and the distribution
$P(\hat{s})$ is known as the Kida distribution \cite{Kida1979,Bouchaud1997,LeDoussal2009}, 
\begin{equation}
\label{eq:Ph2JumpSizesSR}
P(\hat{s}) = \frac{1}{2} \hat{s} e^{-\frac{\hat{s}^2}{4}}\ .
\end{equation}
The scale $\mu$ is related to the density of shocks $\rho_{0}$ through \cite{LeDoussal2009}
\begin{equation}
1 = \rho_{0} \left< x_{2}-x_{1} \right>=\rho_{0} \frac{\mu}{m} \int_0^{\infty}\rmd \hat{s}\, P(\hat{s}) \hat{s} = \rho_{0} \frac{\mu}{m} \sqrt{\pi}\ .
\end{equation}
We thus obtain the scaling form 
\begin{eqnarray}
\label{eq:DeltaPh2UnifFinal}
\Delta_{\theta}(w) & = & \beta m^2 \tilde{\Delta}_{\theta } \left(\hat{w} = \beta m \mu w \right)\\
\nonumber
\tilde{\Delta}_{\theta }(x) & = & \int_{0}^{\infty} \rmd \hat{s}\, \frac{\hat{s}^2}{4\sqrt{\pi}} e^{-\frac{\hat{s}^2}{4}} \left[\frac{\hat{s}x}{e^{\hat{s}x}-1}-\log\left(1-e^{-\hat{s}x}\right)\right]\ .
\end{eqnarray}
Observe that the scaling is different compared to the correlator in the strong-interference phase III: The argument of the scaling function is now $\hat{w}=  \beta m\mu  w$ instead of $\hat{w}=  \sqrt{\beta} m w$ in \eqref{eq:GaussianDeltaWW}.

For small $x$, \eqref{eq:DeltaPh2UnifFinal} has the asymptotic form
\begin{equation}
\label{eq:DeltaPh2UnifAsympt}
\tilde{\Delta}_{\theta }(x) = \frac{1}{4} (\gamma_{\mathrm{E}} - 2 \log{x}) + \mathcal{O}(x)
\end{equation}
This logarithmic singularity arises from the $\phi=\pm \pi$ limit of the integral \eqref{eq:CxShockGen}, and is hence caused by shocks where $Z(w^*)=0$.

\begin{figure}
	\centering
		\includegraphics{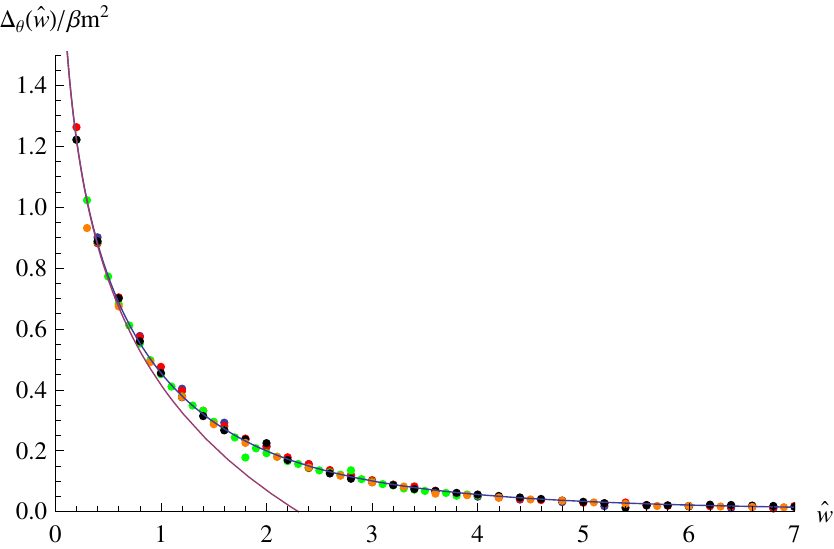}
	\caption{(Color online) Correlator in phase II, with $\theta(x)$ uniform in $[-\pi,\pi]$. Dots: simulation results (blue: $m=0.01$, $\beta=10$, red: $m=0.01$, $\beta=20$, green: $m=0.05$, $\beta=20$, black: $m=0.05$, $\beta=40$, orange: $m=0.05$, $\beta=60$), rescaled as in \eqref{eq:DeltaPh2UnifFinal} with $\mu=0.58$ (for all curves). Blue line: \eqref{eq:DeltaPh2UnifFinal}, red line: asymptotics \eqref{eq:DeltaPh2UnifAsympt}.}
	\label{fig:CxCorrPh2}
\end{figure}

The integral \eqref{eq:DeltaPh2UnifFinal} can be computed numerically
and compared to simulations. We obtain a very good agreement with our
numerical results (see figure \ref{fig:CxCorrPh2}) for various values
of $m$ and $\beta$, providing a non-trivial check for the scaling in
\eqref{eq:DeltaPh2UnifFinal}. The scale $\mu$ is fitted as  $\mu=0.58$,
independent of $m$ or $\beta$ in the  
considered range \footnote{Actually, for the short-range random potential on a discrete lattice
considered here, $\mu$ contains corrections, which are scaling logarithmically with $m$;  see \cite{LeDoussal2009} for more details. If we were to perform the simulations with $m$ varying over several decades, $\mu$ would have to be adjusted correspondingly.}.

\subsection{Example 2: Wrapped gaussian distribution in a short-range potential\label{sec:Ph2Ex2}}
\begin{figure}
	\centering
		\includegraphics{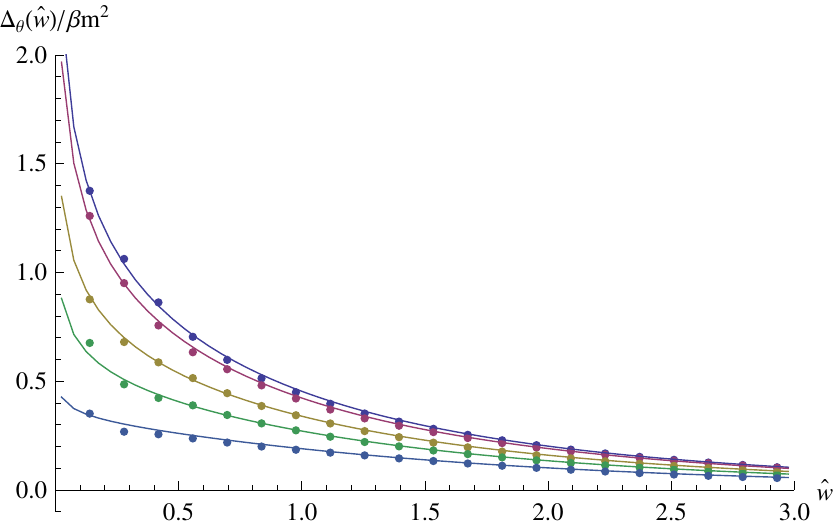}
	\caption{Correlator in phase 2, with $\theta(x)$ wrapped gaussian as in \eqref{eq:WrapGaussAngle}. Dots: simulations (From top to bottom the variance decreases as: $\sigma=2$, $\sigma=1.6$, $\sigma=1.2$, $\sigma=1$, $\sigma=0.8$, the mass is $m=0.01$, and $\beta=120$. Correspondingly, the probability of a jump through zero decreases as $P(\phi=\pm\pi)=0.15,0.13,0.08,0.05,0.01$.), rescaled as in \eqref{eq:DeltaPh2UnifFinal} with $\mu=0.59$ (for all curves). Lines: numerical integration of \eqref{eq:CxShockGen}.}
	\label{fig:CxCorrPh2Wrapped}
\end{figure}

It is  interesting to consider an example where the distribution for $\phi$ is non-uniform.
We take again the phase angle $\theta(x)$ to be uncorrelated at different points. At each point, we assume the distribution of $\theta\in [-\pi;\pi]$ to be a wrapped gaussian distribution with variance $\sigma^2$:
\begin{eqnarray}
\label{eq:WrapGaussAngle}
\tilde P(\theta) & = & \sqrt{\frac{1}{2\pi\sigma^2}}\sum_{n=-\infty}^{\infty} e^{-\frac{1}{2\sigma^2}(\theta+2\pi n)^2} \\
\nonumber
& = & \sqrt{\frac{1}{2\pi\sigma^2}} e^{-\frac{1}{2\sigma^2}\theta^2} \vartheta\left(\frac{\theta}{\sigma^2}i; \frac{2\pi i}{\sigma^2} \right)
\end{eqnarray}
 $\vartheta$ denotes the Jacobi theta function. Note that $\tilde P(\theta)$ is periodic, $\tilde P(\theta+2\pi)=\tilde P(\theta)$. From \eqref{eq:WrapGaussAngle}, the distribution of phase jumps $\phi = \theta_2-\theta_1$ is
\begin{equation}
\label{eq:WrapGaussJump}
P(\phi)=\int_{-\pi}^\pi \tilde P(\theta)\tilde P(\theta+\phi) \, \rmd \theta\ .
\end{equation}
For the random potential, we still assume a short-range random potential as in section \ref{sec:Ph2Ex1}. The distribution of jump sizes $\hat{s}$ is thus still given by \eqref{eq:Ph2JumpSizesSR}. This allows to obtain the full disorder correlator $\Delta_\theta$ by computing the integral \eqref{eq:CxShockGen} numerically. The results in figure \ref{fig:CxCorrPh2Wrapped} compare well to numerical simulations.

One again observes a distinctive logarithmic singularity at $w=0$. This  arises from the $\phi=\pm\pi$ limit of the integral \eqref{eq:CxShockGen}. More precisely, 
\begin{eqnarray}
\nonumber
& & \int_{-\pi}^\pi \rmd \phi\, P(\phi) \sin^2(\phi) \frac{\phi \cot \phi - \frac{sw}{2} \coth \frac{sw}{2}}{\cos 2\phi - \cosh sw}  \\
 & & = - \pi P(\phi=\pm\pi) \log{w} + \mathcal{O}(w^0)
\end{eqnarray}
The integral over $\hat{s}$ is  normalized since
$\int_0^{\infty} \frac{\hat{s}^2}{2\sqrt{\pi}}
e^{-\hat{s}^2/4}\, \rmd \hat{s}=1$, and hence
\begin{equation}
\label{eq:DeltaPh2WrappedAsympt}
\tilde{\Delta}_{\theta }(x)= - \pi P(\phi=\pm\pi) \log{x} + \mathcal{O}(x^0)\ .
\end{equation}
For a uniform distribution of $\theta$,
$P(\phi=\pm\pi)=\frac{1}{2\pi}$ and we recover the $\log$ part of the
result \eqref{eq:DeltaPh2UnifAsympt}.  The constant coefficient of order $w^0$ is  harder to obtain.

In general, the coefficient of the logarithmic singularity at $w=0$ is
proportional to the probability of phase jumps by an angle of
$\phi=\pm\pi$. Thus  intuitively this singularity is caused by shocks
between minima of $V(x)$, where $Z (w^*-0^+)$ changes to $Z (w^*+0^+)
= - k Z (w^*-0^+)$ where $k$ is a
positive number. Note that at any temperature $T>0$, i.e.\ $\beta
<\infty$ the function $Z (w)$ is smooth, thus passes
through zero in our two-well approximation.  This means that the 
Burgers velocity profile has a pole.
Observe that according to \eqref{eq:DeltaPh2WrappedAsympt}, the
logarithmic singularity is present as soon as there is a finite
probability of jumps with an angle of $\phi=\pm\pi$, however small it
may be. This
 shows that in our model, there is no ``sign phase transition'' in the frozen, or pinned phase. This is in agreement with a recent result for a higher-dimensional model \cite{Kim2010} where only $\theta=0$ and $\theta=\pi$, i.e.\ plus and minus signs were considered.

It is straightforward to repeat the analysis above with long-range correlated random potentials $V(x)$. For example, in the case
of the Sinai model explicit expressions for the probability distribution \eqref{eq:Ph2JumpSizesSR} for the jump sizes $s$
are known (see \cite{LeDoussal2009})
but lead to complicated integrals. 

Note that the two-well model is only expected to be valid asymptotically for $\beta \to \infty$. At low but non-zero temperature, we expect subdominant contributions from higher-lying minima which may provide additional
rounding of the singularities discussed above.

In the following, we shall see that the behaviour in the high-temperature phase is quite different.

\section{The high-temperature phase (phase I)\label{sec:Ph1}}
For completeness, we also discuss the disorder correlators in
the high-temperature phase.  In this phase  $Z$ is dominated by the average $\overline{Z}$, and fluctuations are subdominant. As a consequence, for example the quenched average of the free energy is equal to the annealed average of the free energy.

In our one-dimensional model, this phase occurs for sufficiently weak random potentials (for example, short-range correlated $V(x)$ below a  critical value of $\beta$, which increases with system size) and sufficiently weak random-phase disorder (for example, short-range correlated $\theta(x)$ with  finite variance, e.g.\ a wrapped  Gaussian distribution).

To compute the leading-order term for the correlators, let us take the example of short-range real and imaginary disorder, with
\begin{eqnarray}
\overline{V (x)} &=& \overline{\theta (x)} = 0\\ 
\overline{V(x)V(x')} &=& \sigma_V \delta(x-x') \label{62} \\
\overline{\theta(x)\theta(x')} &=& \sigma_\theta \delta(x-x')
\end{eqnarray}
For small $\sigma_V$ and $\sigma_\theta$, we can expand the partition sum in powers of $V$ and $\theta$:
\begin{align}\label{}
&Z(w)  \\
&=\sqrt{\frac{\beta m^2}{2 \pi}} \int_{-\infty}^{\infty} \rmd x\,
\left[1-\beta V(x)- i\theta(x) + ...
\right] e^{-\beta\frac{m^2}{2}(x-w)^2} \nn
\end{align}
The leading order for the effective potential thus becomes:
\begin{eqnarray}
\hat{V}(w) & = & \sqrt{\frac{\beta m^2}{2 \pi}} \int_{-\infty}^{\infty} V(x) e^{-\beta\frac{m^2}{2}(x-w)^2} \, \rmd x\qquad \\
\hat{\theta}(w) & = & \sqrt{\frac{\beta m^2}{2 \pi}}
\int_{-\infty}^{\infty} \theta(x) e^{-\beta\frac{m^2}{2}(x-w)^2}\, \rmd x
\end{eqnarray}
From this, we obtain the leading order for the disorder correlators in
the high-temperature phase: 
\begin{eqnarray}
\label{eq:HighTCorr}
\Delta_{V} (w) & = & \sigma_V \frac{(\beta m^2)^{\frac{3}{2}}}{8\sqrt{\pi}}  \left(2-\hat{w}^{2}  \right)e^{-\frac{\hat{w}^2}{4}} \\ 
\Delta_{\theta} (w) & = & \sigma_\theta \frac{(\beta
m^2)^{\frac{3}{2}}}{8\sqrt{\pi}}  \left(2-\hat{w}^{2}  \right) e^{-\frac{\hat{w}^2}{4}}
\end{eqnarray}
Here, $\hat{w}=m\sqrt{\beta}w$. 

Another way to understand these correlators is through the so-called \textit{exact renormalization group} equations following \cite{LeDoussal2009}.
From \eqref{eq:DefEffPot} and \eqref{eq:IntroZ}, we obtain a flow equation of the form:
\begin{equation}
-m\partial_m \hat{V}(w) = \frac{1}{\beta m^2}\partial_w^2 \hat{V}(w)-\frac{1}{m^2}\left(\partial_w \hat{V}(w)\right)^2
\end{equation}
For the correlator $R(w-w'):=\overline{\hat{V}(w)\hat{V}(w')}$, this gives:
\begin{equation}\label{flowR}
-m\partial_m R(w) = \frac{2}{\beta m^2}\partial_w^2 R(w) +\frac{2}{m^2} S_{110} (0,0,w)
\end{equation}
Here,
$S(w_1,w_2,w_3):=\overline{\hat{V}(w_1)\hat{V}(w_2)\hat{V}(w_3)}$ is
the third cumulant and the subscript $S_{110}$ indicates derivatives
with respect to the first two arguments (notations as in
\cite{LeDoussal2009}). Without the non-linear term,  equation
\eqref{flowR} is the same as  \eqref{eq:FlowTildeF}, solved by
\eqref{eq:TildeFAsympt} with initial conditions \eqref{62}, i.e.
\begin{equation}\label{Rw}
R (w) = \sigma_{V} \sqrt{\frac{\beta m^{2}}{4\pi}} e^{-\frac{\hat{w}^{2}}4 }
\end{equation}
The feeding term for $S$ is of order $R R\sim \beta$, thus subdominant in
$\beta$ for high $T$, and \eqref{Rw} is the complete
solution. Taking two derivatives one obtains $\Delta_{V} (w) =
-\partial_{w}^{2} R (w)$ in agreement with \eqref{eq:HighTCorr}.

\begin{figure}
	\centering
		\includegraphics{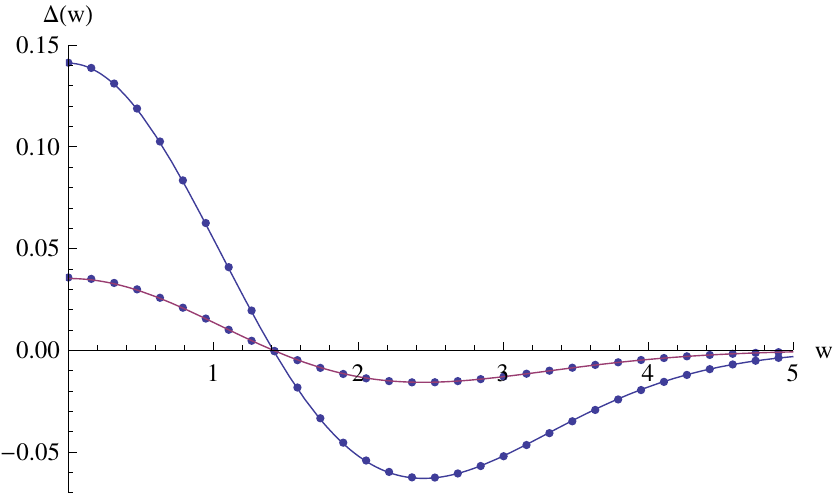}
	\caption{(Color online) Renormalized disorder correlators in the high-temperature phase. Dots: Simulation results for $\beta=0.1, \sigma_V = 1$, $\sigma_\theta = 0.25$, $m=0.5$. Blue (red) line: $\Delta_{V}$ ($\Delta_{\theta}$), as obtained from \eqref{eq:HighTCorr}. No fit parameter.}
	\label{fig:CxCorrHT}
\end{figure}

These results can be compared to simulations in the high-temperature
region. As can be seen on figure \ref{fig:CxCorrHT}, they show excellent agreement.

We thus observe that the behaviour of the model when random phases are
added is very different in the high-temperature phase from what was
observed in the frozen phase (section \ref{sec:Ph2}). For small random
phase disorder, i.e.\ small $\sigma_\theta$, the disorder correlators
stay regular at zero and do not develop any cusp or logarithmic singularity. An intuitive, physical explanation for this can be given:
In the high-temperature phase, the fluctuations of $Z$ are subdominant
compared to the average $\overline{Z}$ in the $m\rightarrow 0$ limit. Hence, even
if there is a finite probability for fluctuations with opposite
phases, a macroscopic number of them would need to occur
simulatenously in order to cancel the average $\overline{Z}$ and lead
to a zero of $Z(w)$. This becomes infinitely improbable in the
$m\rightarrow 0$ limit. On the other hand, in the frozen phase, $Z$
can be approximated by a two-well picture, even in the $m\rightarrow
0$ limit. Then  there is a finite probability for the two minima to
have opposite phases, and thus a finite probability for a shock where $Z$ passes through zero.

\section{Summary and conclusion}
In this paper, we have discussed interference effects in toy models of disordered systems. We considered one-dimensional models, where interference is included through a random complex phase on each lattice site. 

We have obtained the scaling behaviour and asymptotic analytic expressions for the effective disorder correlators in the three phases of the model. For high temperatures small random phases do not change the physics, but strong random-phase disorder leads to a new strong-interference phase. This phase is characterized by a Gaussian distribution of $Z$ centered around zero, and hence a finite density of zeroes of $Z$. For low temperatures, the system is frozen. Introducing random phases changes the structure of the shocks which are observed when a particle is ``dragged'' through the random potential. There, also, zeroes of $Z$ or, equivalently, poles of the complex Burgers velocity field can occur. This physical characterization is seen in the effective disorder correlators as a logarithmic singularity around zero.

A few directions in which the present discussion could be continued come to mind. The most important aspect would be, certainly, to relate the physics observed here to higher-dimensional, more realistic models of interfering quantum systems. In principle, one should be able to obtain the effective disorder correlators from field theory, in the frozen phase e.g. from functional renormalization group methods \cite{LeDoussal2004}. The main technical difficulty, as apparent from our toy model and a preliminary study \cite{Dobrinevski2010} is the behaviour at zero: Instead of a rounding of the linear cusp at finite temperature, we may see a logarithmic singularity. This makes the derivation of a field theory for the frozen phase in the presence of random phases a challenging problem. 

Another direction, which would be interesting to understand better, is the relationship of the present results on the abundance of poles of the Burgers velocity profile to the pole expansion method for the solution of the Burgers equation \cite{Senouf1997,Senouf1996,Bessis1990,Bessis1984}, and the pole condensation phenomena observed in \cite{Bessis1984}.

\begin{acknowledgments}
We thank Jean Dalibard for providing reference \cite{Ovchinnikov1999}, and M. Ortu\~{n}o and A. M. Somoza for valuable discussions. This work was supported by the ANR grant 09-BLAN-0097-01/2 and by the National Science Foundation under Grant No.\ NSF PHY05-51164. We thank the KITP, where some of this work was performed, for hospitality.
\end{acknowledgments}

\medskip

{\it Note added:}

After completion of this paper we became aware of a very recent preprint \cite{Gredat2011} by Gredat, Dornier and Luck which also treats imaginary Brownian
disorder motivated by a connection to reaction-diffusion processes. While the focus is different, i.e. they study the so-called Kesten variable which amounts to
a linear potential regularization, while we study a quadratic well, there is agreement 
whenever the results can be compared.

\appendix
\section{Moments of the partition sum with finite system size $L$ and long-range correlated disorder\label{sec:AppLMoments}}
Let us consider \eqref{eq:CxPartSumL} with $V(x)=0$ and $\theta(x)$ as
defined in \eqref{eq:LRWDef}. In this appendix, we  calculate
explicitely  moments of $Z_{L}$. We  also discuss how they can be
organized to extract the dominant contributions for large $L$.

First, we need to make some technical remarks. Consider the integral
\begin{eqnarray}
\label{eq:GeneralInt}
\lefteqn{I_n^{L}(\lambda_1,...,\lambda_n):=\int_0^L \rmd x_1 \dots \int_0^{x_{n-1}} \rmd x_n\, e^{\sum_{i=1}^n \lambda_i x_i}}&&\\
\nonumber&=&\sum_{k=0}^n e^{\sum_{j=1}^k \lambda_j L} (-1)^{n-k} \prod_{l=1}^k\frac{1}{\sum_{j=l}^k \lambda_j}\prod_{l=k+1}^n\frac{1}{\sum_{j=k+1}^l \lambda_j}
\end{eqnarray}
By introducing the partial sums $\mu_k := \sum_{j=1}^k \lambda_j$, the formula \eqref{eq:GeneralInt} can be rewritten as
\begin{equation}
\label{eq:LaplExpInt}
I_n^{L}(\lambda_j) = \sum_{k=0}^n e^{\mu_k L} \prod_{\stackrel{l=0}{l\neq k}}^n \frac{1}{\mu_k-\mu_l}
\end{equation}
Taking a Laplace transform with respect to $L$, this is further simplified to
\begin{eqnarray}
\nonumber
\text{LT}\{I_n(\lambda_j)\}(s)&=&\int_{L=0}^\infty e^{-s L} I_n^{L} (\lambda_j)\,\rmd L\nn \\
& =& I_{n+1}^{\infty}(-s,\lambda_1,\dots ,\lambda_n)\nn  \\
\label{eq:GeneralIntLT}
&=& \prod_{l=0}^n \frac{1}{s-\mu_l}
\end{eqnarray}
Let us now return to the moments of $ZZ^*$. We would like to evaluate
\begin{eqnarray}
\nonumber\overline{\left(ZZ^*\right)^n}&=&\int_0^L \rmd x_1\dotsb \int_0^L \rmd x_n \int_0^L \rmd y_1\dotsb \int_0^L \rmd y_n\, \\
\nonumber& &  \quad  e^{-\frac{\sigma}{4}\left(\sum_{i,j=1}^n{|x_i-y_j| + |y_i-x_j| - |x_i-x_j| - |y_i-y_j|}\right)}
\end{eqnarray}
If we assume an ordering of the $2n$ variables $x_j$ and $y_j$, the
exponent is a linear combination of these variables. Hence, it is of the form of the integral \eqref{eq:GeneralInt}.
Each ordering of the $x$'s and $y$'s can be mapped bijectively to a
directed path from the lower left to the upper right corner in a
$n\times n$ lattice: The choice of an $x$ corresponds to going
up, the choice of a $y$ to going right\footnote{This generalizes straightforwardly
to general moments like $Z^n \left(Z^*\right)^m$, which give directed paths on an $n\times m$ lattice.}.

Each ordering $\Sigma$ can be defined by a vector ${\Sigma}^x$ with $2n$ entries given by 
\begin{equation}
\Sigma^x_j=\left\{ 
\begin{array}{ccc}
	1 & \mbox{ if $j$th variable is } x \\
	0 & \mbox{ if $j$th variable is }  y \\
\end{array} \right.\ ,
\end{equation}
or equivalently by a vector ${\Sigma}^y$ with $2n$ entries
\begin{equation}
\Sigma^y_j=\left\{ 
\begin{array}{ccc}
	1 &\mbox{ if $j$th variable is } y \\
	0 &\mbox{ if $j$th variable is } x \\
\end{array} \right.\ .
\end{equation}
Then, the resulting values of the $\lambda_j$ for the ordering
$\Sigma$ in the definition \eqref{eq:GeneralInt}  are
\begin{equation}
\frac{2}{\sigma}\lambda_j^\Sigma = (-1)^{\Sigma^x_j}\left[ 2\left(\sum_{l=1}^j \Sigma^x_j - \sum_{l=1}^{j-1}\Sigma^y_j\right) - 1\right]
\end{equation}
A few examples for  $n=2$ and $n=3$  are given in table \ref{tbl:CxOrderingLambdas}.
\begin{table}%
\begin{tabular}{ccc}
Ordering & $\frac{2}{\sigma}\lambda_j$ & $\frac{2}{\sigma}\mu_j$ \\
xyxy & (-1,1,-1,1) & (0,-1,0,-1,0) \\
xyyx & (-1,1,-1,1) & (0,-1,0,-1,0) \\
xxyy & (-1,-3,3,1) & (0,-1,-4,-1,0) \\
xyxyxy & (-1,1,-1,1,-1,1) & (0,-1,0,-1,0,-1,0) \\
xyxxyy & (-1,1,-1,-3,3,1) & (0,-1,0,-1,-4,-1,0) \\
xxxyyy & (-1,-3,-5,5,3,1) & (0,-1,-4,-9,-4,-1,0) \\
\end{tabular}
\caption{The $\lambda_{j}$ and $\mu_{j}$ for some orderings $\Sigma$.}
\label{tbl:CxOrderingLambdas}
\end{table}
In order to apply \eqref{eq:LaplExpInt}, we now need the partial sums $\mu_k$:
\begin{eqnarray}
\nonumber
\frac{2}{\sigma}\mu_k^\Sigma &:=& \frac{2}{\sigma}\sum_{j=1}^k \lambda_j = \sum_{j=1}^k (-1)^{\Sigma^x_j}\left[ 2\left(\sum_{l=1}^j \Sigma^x_j - \sum_{l=1}^{j-1}\Sigma^y_j\right) - 1\right] \\
\label{eq:ComplexIntMu}
&=& -\left[\sum_{l=1}^k\left( \Sigma^x_j -\Sigma^y_j \right)\right]^2
\end{eqnarray}
Again, see table \ref{tbl:CxOrderingLambdas} for a few examples.

Using  formula \eqref{eq:GeneralIntLT}, the the Laplace transform of
the moments  can  be written as:
\begin{equation}
\overline{\left(ZZ^*\right)^n} = (n!)^2 \sum_{\Sigma} I_{2n} (\lambda_j^\Sigma) = (n!)^2 \sum_{\Sigma} \prod_{l=0}^{2n} \frac{1}{s-\mu_l^\Sigma}
\end{equation} 
In the interpretation of $\Sigma$ as a directed path $\Gamma=(w_0,\dotsc ,w_{2n})$, with $w_j$ on the square $n\times n$ lattice and $w_0=(0,0)$, $w_{2n}=(n,n)$, the formula \eqref{eq:ComplexIntMu} obtains a direct interpretation:
$\frac{2}{\sigma}\mu_k^\Sigma$ is  $-d^2$, with $d$ the distance to the diagonal. We thus obtain the interesting formula (setting $\sigma=2$ for simplicity in the following):
\begin{equation}
\label{eq:AppHigherMoments}
\overline{\left(ZZ^*\right)^n} = (n!)^2 \sum_{\stackrel{\Gamma}{\text{path }(0,0)\rightarrow (n,n)}} \prod_{w\in \Gamma} \frac{1}{s+d_w^2}
\end{equation} 
with $d_w$ the distance of $w$ to the diagonal.

For the inverse Laplace transform, no closed formula is evident. However, from \eqref{eq:AppHigherMoments} we can observe the following:
\begin{itemize}
	\item The Laplace transform of $\left(ZZ^*\right)^n$ has poles
	at $s=0$, $s=-1$, $s=-4$, $s=-9$, etc. Hence,
	$\left(ZZ^*\right)^n$ as a function of $L$ can be written as a
	sum of terms of order  $ 1$, $ e^{-L}$, $ e^{-4L}$, $ e^{-9L}$,
	etc.

	\item For large system sizes, the terms suppressed exponentially with $L$ are irrelevant, and hence only the pole at $s=0$ needs to be discussed.
	\item For each path, the pole at $s=0$ has  the form $\frac{1}{s^{z+1}}$ where $z$ is the number of crossings of the diagonal. Its Laplace transform yields $\frac{L^z}{z!}$. Hence, the dominant term for large $L$ is given by the paths with the maximum number of diagonal crossings. 
	\item These are exactly the paths where the $x_{i}$ and $y_{i}$ are  paired, i.e.\ xyxyxyxy... or xyyxxyxy... etc. There are $2^n$ such configurations.
\end{itemize}
The final result is 
\begin{equation}
\nonumber
\overline{\left(ZZ^*\right)^n} = n! (2L)^n +\mathcal{O}(L^{n-1}) + \mathcal{O}(e^{-L})
\end{equation}
This argument provides a somewhat more detailed explanation of why the only configurations contributing to moments of the form $\left(ZZ^*\right)^n$ are those where the replica are pairwise bound. When regularising the system by a harmonic well with mass $m$, we expect similar results, with -- morally speaking -- $L$ replaced by $\frac{1}{m}$. However, we have not found a way to perform a more detailed computation using the regularization with a mass.


\tableofcontents

\end{document}